\documentclass[aps,prd,preprint,superscriptaddress,eqsecnum,nofootinbib]{revtex4}
\usepackage{graphics,epsfig}

\begin{document}
\preprint{WM-01-111}
\preprint{BU-01-22}
%
\title{\vspace*{0.3in}
U(2)-like Flavor Symmetries and Approximate Bimaximal Neutrino Mixing
\vskip 0.1in}
\author{Alfredo Aranda}
\email[]{aranda@BUPHY.bu.edu}
\affiliation{Department of Physics, Boston University, 
590 Commonwealth Ave, Boston, MA 02215}
\author{Christopher D. Carone}
\email[]{carone@physics.wm.edu}
\author{Patrick Meade}
\email[]{prmead@mail.wm.edu}
\affiliation{Nuclear and Particle Theory Group, Department of
Physics, College of William and Mary, Williamsburg, VA 23187-8795}
\date{September, 2001}
\begin{abstract}
Models involving a U(2) flavor symmetry, or any of a number of its 
non-Abelian discrete subgroups, can explain the observed hierarchy 
of charged fermion masses and CKM angles. It is known that a large
neutrino mixing angle connecting second and third generation fields
may arise via the seesaw mechanism in these models, without a fine 
tuning of parameters.  Here we show that it is possible to obtain
approximate bimaximal mixing in a class of models with U(2)-like Yukawa 
textures.  We find a minimal form for Dirac and Majorana neutrino mass
matrices that leads to two large mixing angles, and show that our 
result can quantitatively explain  atmospheric neutrino oscillations  
while accommodating the favored, large angle MSW solution to the solar 
neutrino problem.  We demonstrate that these textures can arise in
models by presenting a number of explicit examples.
\end{abstract}

\maketitle
\section{Introduction}\label{sec:intro}

New data on neutrino oscillations from experiments like 
SuperKamiokande~\cite{sk} and SNO~\cite{sno}, have provided a means of 
testing theories of fermion masses.  The simple idea that the observed 
hierarchies in the quark and charged lepton mass spectrum may be due to 
the sequential breaking of a horizontal symmetry has led to an expansive 
literature on possible symmetries and symmetry breaking 
patterns~\cite{abelian,nonabelian}. Models based on non-Abelian 
flavor symmetries like U(2) are interesting in that Yukawa matrices 
decompose into a small set of flavor group representations, and the 
textures possible after symmetry breaking are often highly restricted. 
Hierarchies in these textures are not difficult to obtain, since each stage 
of flavor symmetry breaking is associated with a small dimensionless 
parameter that appears in the low-energy effective Lagrangian (namely, the 
ratio of a vacuum expectation value to the cut off of the effective theory). 
It is much harder to see how the breaking of a non-Abelian symmetry that 
leads to strictly hierarchical quark and charged lepton Yukawa matrices 
can account for the two large mixing angles suggested by the current solar 
and atmospheric neutrino data~\cite{bahcall1,othernu}.  In this paper, 
we will show how this situation can arise naturally in models with 
``U(2)-like" Yukawa textures; we define what we mean by this more explicitly 
below.  Study of ways in which large neutrino mixing angles can arise in 
U(2)-like models is worthwhile since these theories can potentially explain 
all fermion masses and mixing angles in one consistent framework.

Let us briefly review the minimal U(2) model~\cite{u2}, which has been 
described in detail elsewhere in the literature. U(2) is assumed to be a 
global symmetry that acts across the three standard model generations.  
Quarks and leptons are assigned to ${\bf 2}\oplus {\bf 1}$ representations, so 
that in tensor notation, one may represent the three generations of any 
matter field by $F^a+F^3$, where $a$ is a U(2) index, and $F$ is $Q$, $U$, 
$D$, $L$, or $E$.  A set of symmetry breaking fields are introduced 
consisting of $\phi_a$, $S_{ab}$ and $A_{ab}$, where $\phi$ is a U(2) 
doublet, and $S$ $(A)$ is a symmetric (antisymmetric) U(2) triplet 
(singlet).  These fields are assumed to develop the pattern of vacuum 
expectation values (vevs)
\[
\frac{\langle \phi \rangle}{M_f} = \left(
\begin{array}{c}0 \\ \epsilon \end{array}\right),
\,\,\,\,\,\,\,\,\,\,
\frac{\langle S \rangle}{M_f} = \left(
\begin{array}{cc} 0 & 0 \\
0 & \epsilon \end{array} \right), 
\]\begin{equation}
\mbox{  and  } \,\,\, \frac{\langle A \rangle}{M_f} =
\left(\begin{array}{cc} 0 & \epsilon' \\ -\epsilon' & 0 
\end{array} \right),
\end{equation}
which follows from the sequential symmetry breaking
\begin{equation} \label{eq:symbrk}
U(2) \stackrel{\epsilon}{\rightarrow} U(1) 
\stackrel{\epsilon'}{\rightarrow} \mbox{nothing}.
\end{equation}
This leads to the canonical U(2) texture
\begin{equation} \label{eq:u2yd}
Y_D \sim \left(\begin{array}{ccc}
0 & d_1 \epsilon' & 0 \\
-d_1 \epsilon' & d_2 \epsilon & d_3 \epsilon \\
0 & d_4 \epsilon & d_5 \end{array} \right) \xi  \,\,\, ,
\end{equation}
where $\epsilon \approx 0.02$, $\epsilon' \approx 0.004$, and
$d_1\ldots d_5$ are order one coefficients that are also determined by 
fitting to the data~\cite{u2}. The parameter $\xi$ is explained below.  
Here we have displayed $Y_D$ since the up quark Yukawa matrix requires 
additional suppression factors to explain why 
$m_d :: m_s :: m_b = \lambda^4 :: \lambda^2 :: 1$ while
$m_u :: m_c :: m_t = \lambda^8 :: \lambda^4 :: 1$, where $\lambda \approx 0.22$
is the Cabibbo angle.  For example, in SU(5)$\times$U(2) unified models,
combined GUT and flavor symmetries prevent the $S$ and $A$ flavons from
coupling at lowest order in $Y_U$, provided that the $S$ flavon is chosen to 
transform as a {\bf 75} under SU(5).  However, the presence of an SU(5) 
adjoint field $\Sigma$ with $\langle \Sigma \rangle/M_f \approx \epsilon$ 
leads to the viable texture
\begin{equation}\label{eq:u2yu}
Y_U \sim \left(\begin{array}{ccc}
0 & u_1 \epsilon' \epsilon & 0 \\
-u_1 \epsilon' \epsilon & u_2 \epsilon^2 & u_3 \epsilon \\
0 & u_4 \epsilon & u_5 \end{array} \right).
\end{equation}
The ratio $m_b/m_t$ is fixed by hand through the choice of the small 
parameter $\xi$~\cite{u2}.  Note that the additional suppression factor 
multiplying the $S$ and $A$ flavon vevs can also be accommodated in models 
based on discrete subgroups of U(2) without requiring a grand unified 
embedding. In some of these models, $\xi\sim \epsilon$ is a prediction
of the theory~\cite{acl1}.

The feature crucial to the success of the U(2) model is the existence
of a U(1) subgroup that rotates first generation fields by a phase.
Notice that the $\epsilon$ entries in Eq~(\ref{eq:u2yd}) appear in the
most general way consistent with this symmetry, while the $\epsilon'$
entries which break the U(1) appear only in the first row and column.
The fact that the U(1) breaking is accomplished by the antisymmetric
flavon $A$ alone is a dynamical assumption, at least at the level
of the low-energy effective theory.  From this point of view, there 
is nothing wrong with vevs of order $\epsilon' M_f$ arising, for example, 
in the first component of a doublet or symmetrically in the off-diagonal 
components of $S$.  We will define a ``U(2)-like" model as any one whose 
Yukawa matrices decompose into symmetric triplet, doublet, and
antisymmetric singlet representations, and whose nonvanishing Yukawa entries 
are of a size consistent with the U(2) symmetry breaking pattern given 
in Eq.~(\ref{eq:symbrk}). Let us illustrate this definition with a concrete 
example:

The smallest non-Abelian discrete subgroup of U(2) with {\bf 1}, {\bf 2}, 
and {\bf 3} dimensional representations, and with the multiplication rule
${\bf 2} \otimes {\bf 2} \sim {\bf 3} \oplus {\bf 1}$ is the double 
tetrahedral group, $T'$.  Models based on $G_f = T' \times Z_3$, and the 
breaking pattern~\cite{acl1,acl2}
\begin{equation}\label{eq:symbrk2}
T' \times Z_3 \stackrel{\epsilon}{\rightarrow} Z^D_3 
\stackrel{\epsilon'}{\rightarrow} \mbox {nothing},
\end{equation}
can exactly reproduce the Yukawa textures of a U(2) model when matter fields 
are assigned to appropriate one- and two-dimensional representations.  The 
symmetry $Z_3^D$ is a diagonal subgroup that provides the desired 
phase rotation on first generation fields.  Moreover, doublet, triplet and 
nontrivial singlet $T'$ representations can be found that are in one-to-one 
correspondence with the $\phi$, $S$, and $A$ flavons of the original U(2) 
model.  In  Ref~\cite{acl1}, models based on $T'$ symmetry were constructed
with an additional doublet flavon that affected only the neutrino mass matrix 
textures:
\begin{equation}
\frac{\langle \phi_\nu \rangle}{M_f} \sim \left(\begin{array}{c}
\epsilon' \\ \epsilon \end{array}\right).
\end{equation}
The pattern of vevs in $\phi_\nu$ is the most general one 
consistent with Eq.~(\ref{eq:symbrk2}) and leads to the  
neutrino mass matrix textures~\cite{acl1}
\begin{equation}
M_{LR} = \left(\begin{array}{ccc}
0 & l_1 \epsilon' & l_3 \epsilon' \\
-l_1 \epsilon' & l_2 \epsilon & l_3  \epsilon \\
0 & l_4 \epsilon & 0 \end{array} \right) \langle H_U \rangle  \,\,\, ,
\end{equation}
\begin{equation}
M_{RR} = \left(\begin{array}{ccc}
r_4 {\epsilon'}^2 & r_4  \epsilon \epsilon' & \epsilon' \\
r_4  \epsilon \epsilon'& r_3 \epsilon &   \epsilon \\
\epsilon' &  \epsilon & 0 \end{array} \right) \Lambda_R  \,\,\, ,
\end{equation}
where $\Lambda_R$ is the right-handed neutrino mass scale, and
$r_i$, $l_i$ are $O(1)$ coefficients. These textures are U(2)-like in 
that each entry is of a size consistent with the two-stage symmetry 
breaking pattern in Eq.~(\ref{eq:symbrk});  the precise power of 
$\epsilon$ or $\epsilon'$ that appears is determined in this example 
by the details of the $T'$ group theory.  The   left-handed Majorana 
mass matrix follows from the seesaw mechanism
\begin{equation}
M_{LL} \approx M_{LR} M^{-1}_{RR} M^\dagger_{LR}
\end{equation}
and has the form
\begin{equation}\label{eq:mllexample}
M_{LL} \sim \left( \begin{array}{ccc}
(\epsilon'/\epsilon)^2 & \epsilon'/\epsilon & \epsilon'/\epsilon \\
\epsilon'/\epsilon & 1 & 1 \\
\epsilon'/\epsilon & 1 & 1 \end{array}\right)
\frac{\langle H_U \rangle^2 \epsilon}{\Lambda_R}
\end{equation}
As promised, a large 2-3 mixing angle has emerged from initial U(2)-like 
textures without any adjustment of parameters. The 1-2 mixing angle, of 
order $\epsilon'/\epsilon$, is naturally of the same size as the
Cabibbo angle.  By choosing the $O(1)$ coefficients appropriately,
it is possible to numerically enhance this result to obtain 
the smallest mixing angle values consistent with the MSW large mixing 
angle (LMA) allowed parameter region given in Ref.~\cite{bahcall1}.  Such 
solutions were considered in quantitative detail in Ref.~\cite{acl3}.  
However, the fact that the current data appears to prefer relatively large 
mixing angles suggests another possibility: $M_{LL}$ is a perturbation about
a bimaximal mixing texture~\cite{bimax} that appears at lowest order in 
the symmetry breaking parameters.  It is this possibility that we explore 
in the sections that follow.

\section{Approximate Bimaximal Mixing}
 
What is intriguing about the result in Eq.~(\ref{eq:mllexample}) is
that it is superficially of the form suggested by Haba~\cite{haba} for 
achieving bimaximal mixing,
\begin{equation} \label{eq:haba}
M_{LL} \sim \left( \begin{array}{ccc}
\Phi^2 & \Phi & \Phi \\
\Phi   & 1 & 1 \\
\Phi   & 1 & 1 \end{array} \right) M_0 \,\,\, ,
\end{equation}
where $\Phi$ is a small parameter, and $M_0$ a characteristic mass
scale.  The crucial difference is that the $O(1)$ sub-block of the
Haba texture is assumed to be of rank one, up to corrections of 
order $\Phi$ or smaller.   A diagonalization of the largest entries 
leaves a matrix of the form
\begin{equation}\label{eq:pdiag}
M_{LL} \sim \left(
\begin{array}{ccc}
\Phi^2 & \Phi & \Phi \\
\Phi   & \Phi &  0 \\
\Phi   &  0   &  1 \end{array} \right) M_0 \,\,\, ,
\end{equation}
which requires a large 1-2 rotation angle to diagonalize further. We 
now demonstrate that it is possible to achieve the Haba texture via the 
seesaw mechanism in U(2)-like theories.

Our approach is to determine first the minimal number of entries
in $M_{LR}$ and $M_{RR}$ that can reproduce the rank-one sub-block
of Eq.~(\ref{eq:haba}).  We will then perturb about this texture by lifting 
the smallest number of texture zeros that allows for a viable phenomenology.  
The only organizing principle that we retain in this model-independent 
analysis is that entries in the first row and column of $M_{LR}$ and $M_{RR}$ 
must involve the appropriate power of $\epsilon'$ to be consistent with the 
breaking of some subgroup that rotates the fields of the first generation 
by a phase.  

We begin with the observation that the matrices
\begin{equation}\label{eq:mlrmin}
M_{LR} = \left(\begin{array}{ccc} 0 & 0 & 0 \\ -l_1 \epsilon' & 0 & 0 \\
0 & l_3 \epsilon & 0 \end{array} \right)  \langle H_U \rangle \,\,\, 
\end{equation}
and
\begin{equation}\label{eq:mrrmin}
M_{RR} = \left(\begin{array}{ccc} 0 & 0 & r_2\epsilon' \\ 
0 & r_1 \epsilon & r_2 \epsilon \\
r_2 \epsilon' & r_2 \epsilon & 0 \end{array} 
\right) \Lambda_R
\end{equation}
lead via the seesaw to
\begin{equation}\label{eq:mllmin}
M_{LL} = \frac{\epsilon}{r_1} \left( \begin{array}{ccc}
0 & 0 & 0 \\
0 & l_1^2 & l_1 l_3  \\
0 & l_1 l_3 & l_3^2 
\end{array}\right) \frac{\langle H_U \rangle^2}{\Lambda_R} \,\,\, .
\end{equation}
The sub-block has a vanishing determinant, by inspection\footnote{Note that 
the parameterization in Eqs.~(\ref{eq:mlrmin}) and (\ref{eq:mrrmin})
is completely general, given that we have not specified the size of
$\epsilon$ or $\epsilon'$.}. It is clear from Eq.~(\ref{eq:mllmin}) that 
the entries shown in (\ref{eq:mlrmin}) and (\ref{eq:mrrmin}) are a 
minimal choice; if one sets any of the parameters to zero, one either 
renders $M_{RR}$ singular or loses the large $2$-$3$ mixing angle in the 
final result.  Note also that the textures in Eqs.~(\ref{eq:mlrmin}) 
and (\ref{eq:mrrmin}) are consistent with the symmetry breaking in 
Eq~(\ref{eq:symbrk}), but require one fine tuning
to be obtained: The $1$-$2$ block of $M_{LR}$ can arise only by a specific 
linear combination of symmetric and antisymmetric flavon vevs.  In the more 
realistic textures that follow, such a fine tuning will not be required.  
Finally, we point out that certain texture zeros,  in particular the $3$-$3$ 
entries of $M_{LR}$ and $M_{RR}$, do not appear in the simplest formulation 
of U(2) models. However, as we mentioned earlier, such textures do arise in 
models based on discrete subgroups of U(2)~\cite{acl1,acl2}.  In Section~4 
we will see how they can also arise in models with U(2) symmetry and 
additional Abelian factors.

We now seek the minimal modifications of  Eqs.~(\ref{eq:mlrmin}) and 
(\ref{eq:mrrmin}) that provide for a viable phenomenology and are 
theoretically well motivated.  To avoid any fine tuning between irreducible 
symmetric and antisymmetric representations, the first zero that we choose 
to lift is the $1$-$2$ entry of $M_{LR}$. Hence, we consider the modification
\begin{equation}
M'_{LR} = \left(\begin{array}{ccc} 0 & l_1 \epsilon' & 0 
\\ -l_1 \epsilon' & 0 & 0 \\
0 & l_3 \epsilon & 0 \end{array} \right) \langle H_U \rangle
\end{equation}
which leads to
\begin{equation} \label{eq:mlltwo}
M'_{LL} = \frac{\epsilon}{r_1} \left(\begin{array}{ccc} 
l_1^2 (\epsilon'/\epsilon)^2 & l_1^2 (\epsilon'/\epsilon) 
& l_1 l_3 (\epsilon'/\epsilon) 
\\ l_1^2 (\epsilon'/\epsilon) & l_1^2 & l_1 l_3 \\
l_1 l_3 (\epsilon'/\epsilon) & l_1 l_3 & l_3^2 \end{array} \right)
\frac{\langle H_U \rangle^2}{\Lambda_R}.
\end{equation}
If we identify $\epsilon'/\epsilon$ with $\Phi$, we obtain a texture of 
the same form as Eq.~(\ref{eq:haba}), with the desired rank one sub-block.
Unfortunately, the texture in Eq.~(\ref{eq:mlltwo}) is still not viable due 
to a specific relationship between the coefficients: the $1$-$2$ and $1$-$3$ 
entries appear in the same ratio as the $2$-$2$ and $2$-$3$ entries.  
Diagonalization of the largest elements leaves a matrix of the form
\begin{equation}
M_{LL} \sim \left(
\begin{array}{ccc}
\Phi^2 & 0 & \Phi \\
0   &  0 &  0 \\
\Phi   &  0   &  1 \end{array} \right) M_0
\end{equation}
and no large $1$-$2$ mixing angle is obtained. It is necessary to lift
at least one additional texture zero in order to disrupt this proportionality 
of coefficients.  We find that the minimal choice, in which only one additional
entry is altered, is unique:
\begin{equation} \label{eq:minformin}
M_{RR} = \left(\begin{array}{ccc} 0 & 0 & r_2\epsilon' \\ 
0 & r_1 \epsilon & r_2 \epsilon \\
r_2 \epsilon' & r_2 \epsilon & 0 \end{array} 
\right) \Lambda_R ,
\,\,\,\,\,\,\,\,\,\,
M''_{LR} = \left(\begin{array}{ccc} 0 & l_1 \epsilon' & l_2 \epsilon' 
\\ -l_1 \epsilon' & 0 & 0 \\
0 & l_3 \epsilon & 0 \end{array} \right) \langle H_U \rangle \,\,\, .
\end{equation}
From here we finally obtain
\begin{equation}\label{eq:minformout}
M''_{LL} = \frac{\epsilon}{r_1} \left(\begin{array}{ccc} 
l_1^2 (\epsilon'/\epsilon)^2 & (l_1^2-l_1 l_2 r_1/r_2) (\epsilon'/\epsilon) 
& l_1 l_3 (\epsilon'/\epsilon) 
\\ (l_1^2-l_1 l_2 r_1/r_2)(\epsilon'/\epsilon) & l_1^2 & l_1 l_3 \\
l_1 l_3 (\epsilon'/\epsilon) & l_1 l_3 & l_3^2 \end{array} \right)
\frac{\langle H_U \rangle^2}{\Lambda_R}\,\,\, .
\end{equation}
We will refer to Eqs.~(\ref{eq:minformin}) and (\ref{eq:minformout}) 
as our minimal bimaximal mixing textures.  

At this point, it is important that we be specific on the meaning of
the zero entries in Eq.~(\ref{eq:minformout}).  We assume simply that
there are no contributions to these entries at linear order in the
symmetry-breaking parameters.   As we will see in Section~4, most realistic 
models imply that texture zeroes are lifted at some order in the flavor
expansion, unless those entries are protected by holomorphy.  This is of 
significance to the phenomenological analysis presented in the next section 
for the following reason:  While the ratio $\Delta m^2_{23}/\Delta m^2_{12} 
\approx \epsilon^2/{\epsilon'}^2 \approx 25$ that follows from 
Eq.~(\ref{eq:minformout}) is naturally of the right size to account for 
atmospheric and LMA solar neutrino oscillations, the experimentally 
preferred value of $\theta_{12}$ is noticeably less than $45^\circ$.  
Corrections to the zero entries allow us numerically to obtain mixing
angles consistent with allowed 95\% confidence region. In particular,
we will study the more general form
\begin{equation} \label{eq:nonminformin}
M_{RR} = \left(\begin{array}{ccc} 0 & 0 & r_2\epsilon' \\ 
0 & r_1 \epsilon & r_2 \epsilon \\
r_2 \epsilon' & r_2 \epsilon & r_3 \epsilon^2 \end{array} 
\right) \Lambda_R ,
\,\,\,\,\,\,\,\,\,\,
M''_{LR} = \left(\begin{array}{ccc} 0 & l_1 \epsilon' & l_2 \epsilon' 
\\ -l_1 \epsilon' & 0 & 0 \\
0 & l_3 \epsilon & 0 \end{array} \right) \langle H_U \rangle \,\,\, .
\end{equation}
since the higher-order $r_3$ entry is quite effective at allowing
adjustment of $\theta_{12}$, and is easily accommodated in realistic
models.   It is worth pointing out that U(2)-like values for $\epsilon$ 
and $\epsilon'$ are not consistent with the LOW or vacuum oscillation 
solutions to the solar neutrino problem, since each requires a value of 
$\Delta m^2_{23}/\Delta m^2_{12}$ that is much larger than that 
predicted from Eq.~(\ref{eq:minformout}).  

\section{Numerical Analysis} \label{numerical}

We now study the textures in Eq.~(\ref{eq:minformin}) and 
(\ref{eq:nonminformin}) numerically, and show that atmospheric
and large angle solar neutrino oscillations can be obtained.  In the
spirit of model independence, we assume a form for the charged
lepton Yukawa matrix that arises generically in U(2)-like models:
\begin{equation} \label{eq:u2ye}
Y_L \sim \left(\begin{array}{ccc}
0 & c_1 \epsilon' & 0 \\
-c_1 \epsilon' & 3 c_2 \epsilon & c_3 \epsilon \\
0 & c_4 \epsilon & c_5 \end{array} \right) \xi  \,\,\, .
\end{equation}
The factor of $3$ that multiplies $c_2$ is the famous one suggested
by Georgi and Jarlskog~\cite{gj}, and arises as a consequence of
grand unified group theory.  We fit to leptonic observables while
fixing $\epsilon = 0.02$ and $\epsilon^{\prime} = 0.004$; these 
are the preferred values obtained in fitting Eq.~(\ref{eq:u2yd}) and 
Eq.~(\ref{eq:u2yu}) to quark masses and CKM angles. This constrained
fit is a reasonable approximation to a more involved global one, given 
the relatively large experimental uncertainty on each of the neutrino 
observables.

We assume that the textures $M_{LL}$ and $Y_L$ are defined at some
high scale, which we take to be $M_{\rm GUT} \approx 2 \times 10^{16}$ GeV,
and perform a renormalization group analysis of the gauge and Yukawa couplings.
We do this by solving the one-loop renormalization group equations (RGE's)
of the MSSM~\cite{BBO} from $M_{\rm GUT}$  down to the electroweak scale 
taken to be $m_t = 175$ GeV.

Values of the gauge couplings at $M_{\rm GUT}$ are obtained by
starting with the precision values extracted at the scale $M_Z$~\cite{PDG},
\begin{eqnarray}
  \alpha_1^{-1} (M_Z) & = & 59.99  \pm 0.04,   \nonumber \\
  \alpha_2^{-1} (M_Z) & = & 29.57  \pm 0.03,   \nonumber \\
  \alpha_3^{-1} (M_Z) & = & 8.40   \pm 0.13.
\end{eqnarray}
The gauge couplings are run from $M_Z$ to $m_t$ using the one-loop
Standard Model (SM) RGE's, and then from $m_t$ to $M_{\rm GUT}$ using the
one-loop MSSM RGE's. The RGE for the neutrino Majorana mass matrix $M_{LL}$ 
was computed in Ref.~\cite{BLP} and is included here in order to
complete the RGE evolution for all observables. 

In order to perform the fits we incorporate experimental and theoretical 
uncertainties on the observables. For the charged leptons, they are either 
those appearing in Ref.~\cite{PDG} or $1\%$ of the central value of the
given datum, whichever is larger. The latter, theoretical uncertainty takes
into account that two-loop corrections to the running and possible 
high-scale threshold corrections have been neglected. The low-energy 
neutrino observables are taken to be
\begin{eqnarray} \label{oldneut}
  & & 4 < \frac{\Delta m_{23}^2}{\Delta m_{12}^2} < 200, \nonumber \\
  & & \sin^2 2\theta_{23} > 0.88, \nonumber \\
  & & 0.2 < \tan^2 \theta_{12} < 0.9 \,\,\, ,
\end{eqnarray}
which were extracted from Refs.~\cite{bahcall1,superk}. Notice that we only 
need to reproduce the ratio $\Delta m^2_{23}/\Delta m^2_{12}$ since the 
right-handed neutrino scale $\Lambda_R$ is freely adjustable. For the 
sake of having meaningful uncertainties, a parameter whose lower bound is 
much smaller than its upper bound is converted into its logarithm.  Instead 
of Eq.~(\ref{oldneut}), we use
\begin{eqnarray}
  \ln \left( \frac{\Delta m_{23}^2}{\Delta m_{12}^2} \right) & = & 3.34
  \pm 0.98 , \nonumber \\
  \sin^2 2\theta_{23} & = & 0.94 \pm 0.03 , \nonumber \\
  \tan^2 \theta_{12} & = & 0.55 \pm 0.18 .
\end{eqnarray}
In order to determine whether one can find a choice of
parameters (generically denoted by $k_i$) which are $O(1)$ 
and at the same time reproduce the values of observables, we perform 
a $\chi^2$ minimization.  The full analysis consists of choosing initial 
values for the $O(1)$ coefficients $k_i$, for 
fixed\footnote{We work with $\tan\beta=3$.  Qualitatively, our
results are insensitive to changes in $\tan\beta$ of order unity.}
$\tan\beta$, running the RGE's down to $m_t$, and comparing observables 
with their experimental values to compute $\chi^2$. Then, the parameters 
$k_i$ are adjusted and the procedure repeated until a minimum of $\chi^2$ 
is obtained.

Our $\chi^2$ function assumes a somewhat nonstandard form.
Lepton masses and neutrino mixing angles are converted to Yukawa couplings
$y_i^{\rm expt} \pm \Delta y_i$, and contribute an amount
\begin{equation}
  \Delta \chi^2 = \left( \frac{y_i^{\rm expt} - y_i}{\Delta y_i}
  \right)^2
\end{equation}
to $\chi^2$, as usual.  There are 6 observables (3 charged lepton masses,
2 neutrino mixing angles, and 1 neutrino mass ratio) and 11 parameters
$k_i$; on the surface, it seems that the fit is always under-constrained.  
However, our demand that the parameters $k_i$ lie near unity imposes 
additional restrictions, which we include by adding terms to $\chi^2$ of 
the form
\begin{equation}
  \Delta \chi^2 = \left( \frac{\ln |k_i|}{\ln 3} \right)^2
\end{equation}
for each $i$.  Thus, the parameters $k_i$ are effectively no longer
free, but are to be treated analogously to pieces of data, each of
which contributes one unit to $\chi^2$ if it is as large as 3 or as
small as 1/3~\footnote{The choice of $3$ is a matter of taste.}.  
Thus, the value of $\chi^2_{\rm min}$ determining a `good' fit is
6, since there are 6 pieces of true data and effectively no
{\em unconstrained\/} fit parameters.  We find that it is not 
difficult to obtain parameters $k_i$ that work, as one can see in
Tables I and II.

\begin{table}
\begin{tabular}{llllll}
\mbox{Observable \qquad} 
& \mbox{Expt. value \qquad} & \mbox{Fit A \qquad } & 
\mbox{Fit B \qquad} & 
\mbox{Fit C \qquad } & \mbox{Fit D  } \\ \hline\hline

$m_e$ & $0.511 \pm 1\% $ & $0.512$ & $0.511$ & $0.511$  & $0.512$ \\

$m_\mu$ & $106 \pm 1\%$ & $106$ & $106$ & $106$ & $106$ \\

$m_\tau$ & $1777 \pm 1\%$ & $1778$ &  $1777$ & $1778$ & $1777$ \\

$\Delta m_{23}^2 / \Delta m_{12}^2$ & $4$---$200$ & $40$  & $34$ & $35$ 
& $42$ \\

$\ln \left( \Delta m_{23}^2 / \Delta m_{12}^2 \right)$ & $3.34 \pm
0.98$ & $3.7$ & $3.5$ & $3.6$ & $3.7$ \\

$\tan^2 \theta_{12}$ & $0.20$ --- $0.90$ & $0.89$ & $0.66$ & $0.66$ 
& $0.88$ \\

$\sin^2 2\theta_{23}$ & $> 0.88$  & $0.94$ & $0.93$ & $0.94$ & $0.93$ \\

$\sin^2 2\theta_{13}$ & $< 0.1$---$ 0.3$ & $ 0.24$ & $0.01$ & $0.04$ &
$0.24$ \\
\hline
\hline
\end{tabular}
\caption{Experimental values versus fit central values for observables
using the inputs of Table~\ref{tab:coefficients}.  Masses are in MeV and 
all other quantities are dimensionless. Error ranges indicate the larger of 
experimental or 1\% theoretical uncertainties, as described in
the text.}
\label{tab:data}
\end{table}
\begin{table}
    \begin{tabular}{ccccc} 
      Fit & A & B & C & D \\ \hline\hline
      $\chi^2$ & $3.451421$& $2.98341084$ & $3.72175717$  & $8.29507256$   \\
      $c_1$ &  $0.47674$ & $0.476627409$ &  $0.476760209$ & $0.476436228$ \\
      $c_2$ &  $0.46998$ & $-0.465719551$ &  $-0.440337121$ & $0.462245226$ \\
      $c_3$ & $0.99173$ & $0.82251513$  &  $1.39617729$  & $0.9901492$ \\
      $c_4$ &  $1.0226$  &  $0.89786166$  & $0.76331389$ & $0.559998155$ \\
      $c_5$ &  $0.45998$ &  $-0.460312963$ & $0.46006763$ &  $0.460330635$ \\ 
      $l_1$ &  $1.3715$  &  $1.1396178$ & $0.674030304$ &  $0.434771806$ \\
      $l_2$ &  $-0.51276$  &  $1.21720707$ & $1.68183231$ &  $1.90096331$  \\
      $l_3$ &  $1.4191$ &  $1.14355946$ & $0.920410395$ &  $0.600989103$ \\
      $r_1$ &  $1.1785$  &  $1.16333687$ & $1.38230038$ &  $0.481762707$ \\
      $r_2$ &  $0.36925$ &  $0.381390542$ & $0.589898586$ &  $0.280204356$ \\
      $r_3$ &  $2.2979$  &  $-1.69190395$  & $-2.8438561$  &  $0.0$\\
      \hline\hline
    \end{tabular}
  \caption{O(1) coefficients from four representative fits 
    with $\tan \beta = 3.0$. The observables computed using these values 
    are shown in Table~\ref{tab:data}. Fit D correspond to the minimal 
    $r_3 =0$ case.}
  \label{tab:coefficients}
\end{table}

For each fit shown a value of the mixing angle $\theta_{13}$ was 
obtained.  While there is no experimental evidence for $1$-$3$
neutrino oscillations, an eventual positive signal could help to
distinguish between possible models. In Fig. 1, we plot the values 
of $\theta_{13}$ vs. $\chi^2$ for a number of different fits.
Each dot in the figure corresponds to a different set of randomly 
generated initial values for the parameters $k_i$, {\em i.e.} a
different local minimum of the $\chi^2$ function. We compare this to 
the bound $\sin^2 2 \theta_{13} < 0.1 - 0.3$~\cite{king}, which
is indicated in the figure by horizontal lines.  The green dots 
correspond to fits where $\theta_{12}$ is above the 95\% C.L. bound.

\begin{figure}
  \begin{centering}
    \def\epsfsize#1#2{0.5#2}
    \hfil\hspace{-10em} \epsfbox{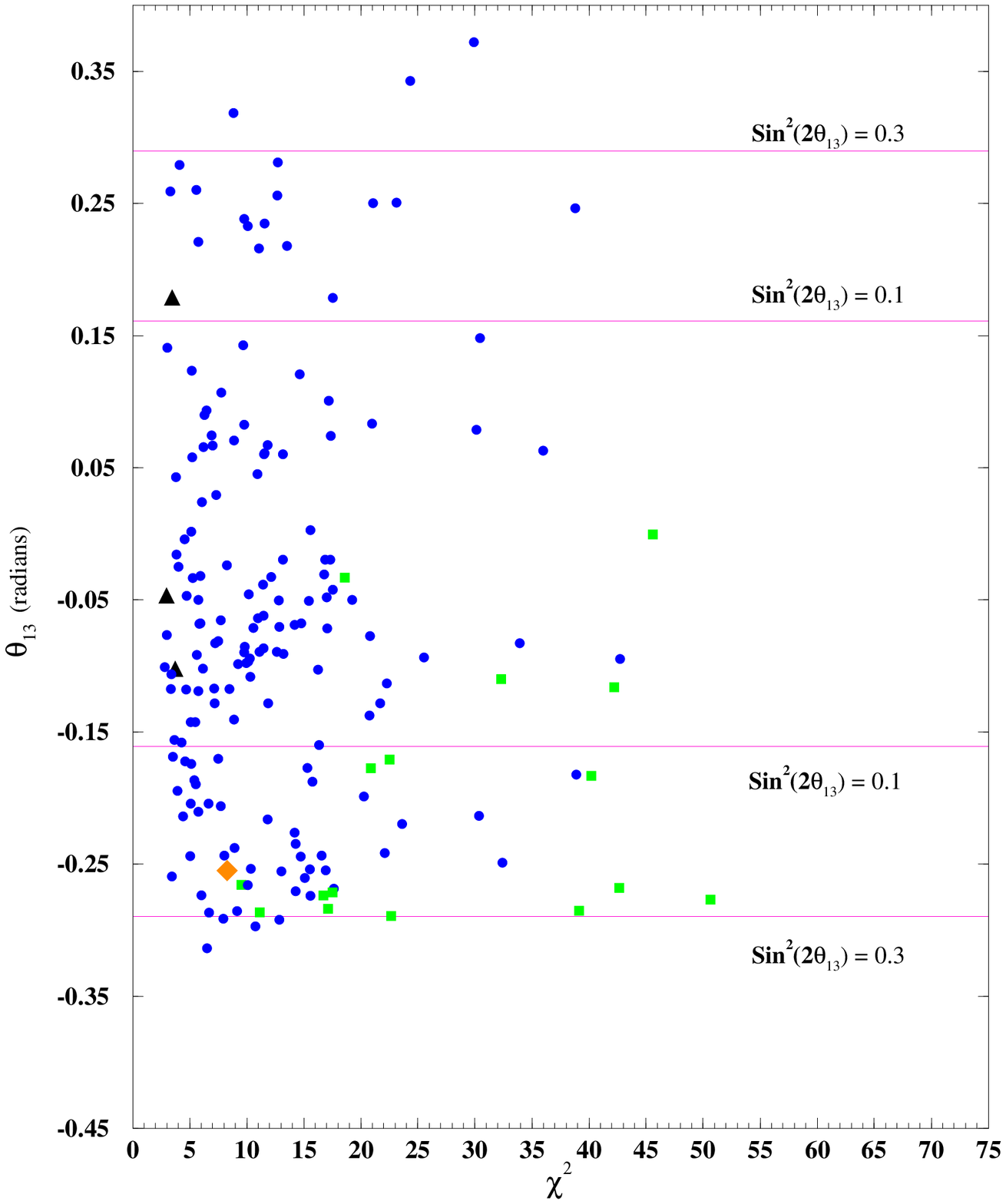} \hfill
    \caption{Values of $\theta_{13}$ vs. $\chi^2$ for $202$ different 
      randomly generated fits. The horizontal lines are the bounds 
      discussed in the text. Blue dots correspond to fits 
      in which all observables are within the desired experimental 95\% C.L. 
      regions for atmospheric and LMA solar neutrino oscillations.  
      Green squares correspond to fits that had a $\theta_{12}$ slightly 
      above the upper bound for the LMA solution. The 3 black 
      triangles correspond to the fits A, B, and C, and the orange diamond 
      is the best fit with $r_3 = 0$, i.e. fit D in the text.}
  \end{centering}
  \label{fig:fig1}
\end{figure}
\begin{figure}
  \begin{centering}
    \def\epsfsize#1#2{0.5#2}
    \hfil\hspace{-10em} \epsfbox{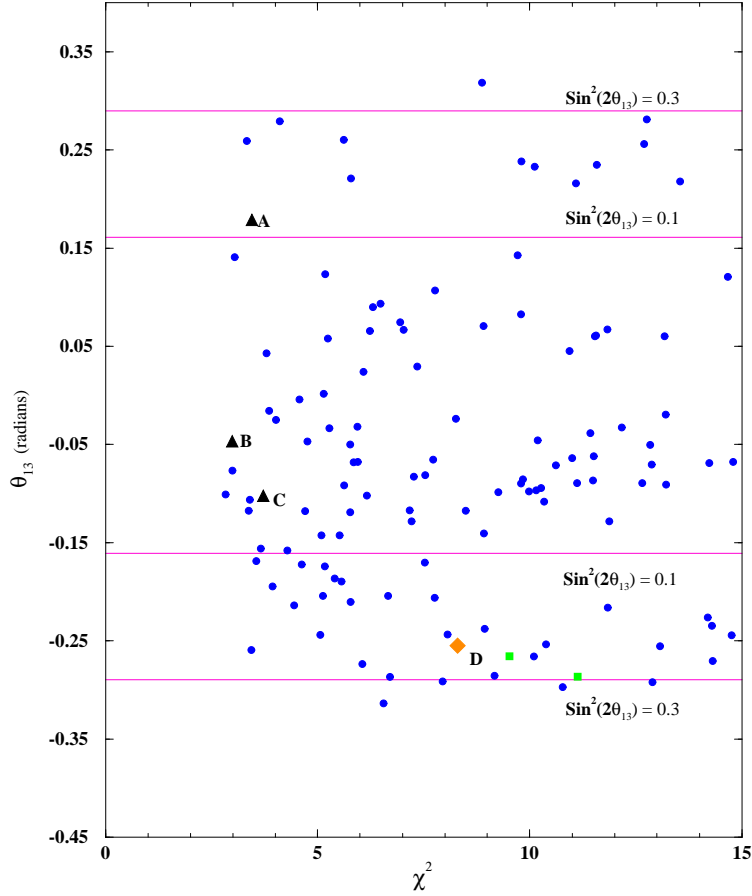} \hfill
    \caption{Here we show the fits with a $\chi^2 < 15$.}
  \end{centering}
  \label{fig:fig2}
\end{figure}

\section{Models}

We now demonstrate that it is possible to construct  models
that realize the textures studied numerically in the previous section.
We aim for the basic forms
\begin{equation} \label{eq:minagain}
M_{RR} = \left(\begin{array}{ccc} 0 & 0 & r_2\epsilon' \\ 
0 & r_1 \epsilon & r_2 \epsilon \\
r_2 \epsilon' & r_2 \epsilon & 0 \end{array} 
\right) \Lambda_R ,
\,\,\,\,\,\,\,\,\,\,
M''_{LR} = \left(\begin{array}{ccc} 0 & l_1 \epsilon' & l_2 \epsilon' 
\\ -l_1 \epsilon' & 0 & 0 \\
0 & l_3 \epsilon & 0 \end{array} \right) \langle H_U \rangle \,\,\, .
\end{equation}
As mentioned earlier, it will require more than just U(2) symmetry to
account for these textures.  For one, there is no invariant $3$-$3$
entry in each matrix, as one would expect in a minimal U(2) model with the
generations assigned to ${\bf 2}+{\bf 1}$ representations.  Moreover, 
these textures imply the existence of three doublet fields, with distinct 
couplings, while U(2) provides only for one.  The simplest approach, which
we will utilize here for the purposes of providing an existence proof, is
to extend the U(2) symmetry by an additional Abelian factor~\cite{u2u1}.
At the end, we describe how similar models may be obtained using
non-Abelian discrete subgroups of U(2). We present examples
that do and do not require a unified gauge group, respectively: 

\subsection{SU(5)$\times$U(2)$\times$$Z_5$}

In this model we let the superfields $Q$, $U$, $D$, $L$, $E$ transform as
\begin{equation}
{\bf{2}}_{0}\oplus {\bf{1}}_{0}
\end{equation}
where the subscript indicates the $Z_{5}$ charge ({\em i.e.} the
subscripts add modulo 5). The Higgs fields $H_{U,D}$ transform as 
trivial singlets under both $U(2)$ and $Z_{5}$.  In addition, there
is a set of  `ordinary' flavons that are $Z_5$ singlets, but transform 
under SU(5) as follows:
\begin{equation}
\begin{array}{c}
S_0\sim 3_{0} \sim \mathbf{75} \\
A_0\sim 1_{0} \sim \mathbf{1} \\
\phi_0 \sim 2_{0} \sim \mathbf{1} \\
\end{array}
\end{equation}
These assignments allow us to reproduce the conventional Yukawa
textures of the unified SU(5)$\times$U(2) model. The desired neutrino 
textures are obtained by introducing right-handed neutrinos transforming 
non-trivially under the
$Z_{5}$ factor:
\begin{equation}
\nu_{R} \sim {\bf 2}_{2} \oplus {\bf 1}_{4}
\end{equation}
The neutrino Dirac and Majorana mass matrices decompose under this
symmetry as
\begin{equation}
M_{LR}\sim\left(\begin{array}{cc}
{\bf \left[3_3\oplus 1_3\right]} & {\bf \left[2_1\right]} \\
{\bf \left[2_3\right]} & {\bf \left[1_1\right]} \end{array}\right), \qquad
M_{RR}\sim\left(\begin{array}{cc}
{\bf \left[3_1\right]} & {\bf \left[2_4\right]} \\
{\bf \left[2_4\right]} & {\bf \left[1_2\right]} \end{array}\right) \,\,\, .
\end{equation}
We introduce the SU(5)-singlet flavons
\[
\frac{\langle A_{3}\rangle}{M_{f}}\sim {\bf 1_{3}} \sim \left(
\begin{array}{cc}
0 & \epsilon' \\
-\epsilon' & 0
\end{array}\right),\qquad
\frac{\langle S_{1}\rangle}{M_{f}}\sim {\bf 3_{1}}\sim
\left(\begin{array}{cc}
0 & 0 \\
0 & \epsilon
\end{array}\right),
\]
\begin{equation}
\frac{\langle\phi_{3}\rangle}{M_{f}} \sim {\bf 2_{3}} \sim
\left(\begin{array}{c}
0 \\
\epsilon \end{array}\right), \qquad
\frac{\langle\phi_{1}\rangle}{M_{f}} \sim {\bf 2_{1}} \sim
\left(\begin{array}{c}
\epsilon' \\
0 \end{array}\right), \qquad
\frac{\langle\phi_{4}\rangle}{M_{f}} \sim {\bf 2_{4}} \sim
\left(\begin{array}{c}
\epsilon' \\
\epsilon \end{array}\right) \,\,\, ,
\end{equation} 
and thus arrive at the correct textures for $M_{LR}$ and $M_{RR}$:
\begin{equation}
M_{LR}\sim\left(\begin{array}{cc}
A_{3} & \phi_{1} \\
\phi_{3} & 0
\end{array}\right)\sim
\left(\begin{array}{ccc} 0 & \epsilon' & 
\epsilon' \\
-\epsilon' & 0 & 0 \\
0 & \epsilon & 0
\end{array}\right)\langle H_{U}\rangle \,\,\, ,
\end{equation}
\begin{equation}
M_{RR}\sim\left(\begin{array}{cc}
S_{1} & \phi_{4} \\
\phi_{4} & 0
\end{array}\right)\sim
\left(\begin{array}{ccc} 0 & 0 & 
\epsilon' \\
0 & \epsilon & \epsilon
 \\
\epsilon' & \epsilon &
0
\end{array}\right)\Lambda_{R} \,\,\, .
\end{equation}
Notice that the $Z_5$ charge assignments of the new flavons prevent
them from affecting the lowest order textures for $Y_U$, $Y_D$ and
$Y_E$.  Thus, the predictions of the minimal unified U(2) model are
maintained.  The pattern of vevs in the doublet flavons is a dynamical 
assumption, at the level of our effective field theory analysis, but 
is at least well motivated:  it is known that minima of potentials occur at 
enhanced symmetry points, and the pattern of vevs is one 
consistent with the sequential breaking in Eq.~(\ref{eq:symbrk}).  
Presumably, an explicit high-energy model would involve a complicated
flavon potential, and different patterns of $\epsilon$ and $\epsilon'$ 
might arise depending on differing parameter choices.  We do not consider
this issue further in this paper.  For some discussion of possible flavon
potentials in U(2) models, see Ref.~\cite{u2pot}.  Finally, we point out
that the presence of additional fields with vacuum expectation
values can perturb these textures; this is not unlikely given that
additional fields are usually required in constructing a realistic flavon
potential.  As an example, the presence of a doublet $\overline{\phi}_4$ 
transforming as ${\bf \overline{2}_4}$ with an $\epsilon$ vev alters none 
of these textures at lowest order. (Unlike SU(2), the ${\bf 2}$ and 
${\bf \overline{2}}$ reps are distinct.)  However, the product 
$\overline{\phi}_4 \phi_3 \sim {\bf 1_2}$ provides the $\epsilon^2$ 
perturbation in the $3$-$3$ entry of $M_{RR}$ considered in the numerical 
analysis.

\subsection{U(2)$\times$U(1)}

Here we show that the inclusion of an additional U(1) symmetry is sufficient 
for constructing viable models, even if there is no unified gauge group.  
Aside from predicting our desired textures Eq.~(\ref{eq:minagain}), we now 
must also account for the additional suppression factor in $Y_U$, discussed 
in the Introduction, that originated previously from the SU(5) transformation 
properties of the flavons. We accomplish this by allowing the charged 
fermions to transform nontrivially under the additional symmetry.  We let
\begin{equation}
\begin{array}{c}
Q,U,E \sim \mathbf{2}_{1}\oplus\mathbf{1}_{0} \\
D,L \sim \mathbf{2}_{1}\oplus\mathbf{1}_{1}
\end{array}
\end{equation}
while the Higgs fields and the right-handed neutrinos transform as
\begin{equation}
\begin{array}{c}
H_{U,D}\sim \mathbf{1}_{0}  \\
\nu_{R}\sim \mathbf{2}_{0}\oplus\mathbf{1}_{3}
\end{array}
\end{equation}
Proceeding as before, the various Yukawa and mass matrices
have the transformation properties:
\[
Y_{U}\sim\left(\begin{array}{cc}
{\bf \left[3_{-2}\oplus 1_{-2}\right]} & {\bf \left[2_{-1}\right]} \\
{\bf \left[2_{-1}\right]} & {\bf \left[1_{0}\right]}
\end{array}\right)\,\,
Y_{D}\sim\left(\begin{array}{cc}
{\bf \left[3_{-2}\oplus 1_{-2}\right]} & {\bf \left[2_{-2}\right]} \\
{\bf \left[2_{-1}\right]} & {\bf \left[1_{-1}\right]}
\end{array}\right) \,\,
Y_{L}\sim\left(\begin{array}{cc}
{\bf \left[3_{-2}\oplus 1_{-2}\right]} & {\bf \left[2_{-1}\right]} \\
{\bf \left[2_{-2}\right]} & {\bf \left[1_{-1}\right]}
\end{array}\right)
\]
\begin{equation}
M_{LR}\sim\left(\begin{array}{cc}
{\bf \left[3_{-1}\oplus 1_{-1}\right]} & {\bf \left[2_{-4}\right]} \\
{\bf \left[2_{-1}\right]} & {\bf \left[1_{-4}\right]}
\end{array}\right) \qquad
M_{RR}\sim\left(\begin{array}{cc}
{\bf \left[3_{0}\right]} & {\bf \left[2_{-3}\right]} \\
{\bf \left[2_{-3}\right]} & {\bf \left[1_{-6}\right]}
\end{array}\right)
\end{equation}
We  introduce the set of flavons
\[
\frac{\langle S_{0}\rangle}{M_{f}}\sim 3_{0}\sim
\left(\begin{array}{cc}
0 & 0 \\
0 & \epsilon
\end{array}\right)\qquad
\frac{\langle A_{-1}\rangle}{M_{f}}\sim 1_{-1}\sim
\left(\begin{array}{cc}
0& \epsilon' \\
-\epsilon' & 0
\end{array} \right)
\]
\begin{equation}
\frac{\langle\phi_{-1}\rangle}{M_{f}} \sim 2_{-1} \sim
\left(\begin{array}{c}
0 \\
\epsilon \end{array}\right)\qquad
\frac{\langle\phi_{-3}\rangle}{M_{f}} \sim 2_{-3} \sim
\left(\begin{array}{c}
\epsilon' \\
\epsilon \end{array} \right) \qquad
\frac{\langle\phi_{-4}\rangle}{M_{f}} \sim 2_{-4} \sim
\left(\begin{array}{c}
\epsilon' \\
0 \end{array}\right)\qquad\] \[
\frac{\langle\chi_{-1}\rangle}{M_{f}}\sim 1_{-1} \sim \epsilon
\end{equation}
from which we obtain 
\begin{equation}
Y_{U}\sim\left(\begin{array}{cc}
\phi^{2}_{-1} + A_{-1}\chi_{-1} & \phi_{-1} \\
\phi_{-1} & 1
\end{array}\right)\sim
\left(\begin{array}{ccc} 0 & \epsilon\epsilon' & 0 \\
-\epsilon\epsilon' & \epsilon^2 & \epsilon \\
0 & \epsilon & 1
\end{array}\right)
\end{equation}
\begin{equation} 
Y_{D}\sim \left(\begin{array}{cc}
\phi^{2}_{-1} + A_{-1}\chi_{-1} & \phi_{-1}\chi_{-1} \\
\phi_{-1} & \chi_{-1}
\end{array}\right)\sim
\left(\begin{array}{ccc} 0 & \epsilon' & 0 \\
-\epsilon' & \epsilon & \epsilon \\
0 & 1 & 1
\end{array}\right)\epsilon
\end{equation}
\begin{equation}
Y_{L}\sim\left(\begin{array}{cc}
\phi^{2}_{-1} + A_{-1}\chi_{-1} & \phi_{-1} \\
\phi_{-1}\chi_{-1} & \chi_{-1}
\end{array}\right)\sim
\left( \begin{array}{ccc} 0 & \epsilon' & 0 \\
-\epsilon' & \epsilon & 1 \\
0 & \epsilon & 1
\end{array}\right) \epsilon  \,\,\, ,
\end{equation}
as well as the  neutrino Dirac and Majorana mass matrices
\begin{equation}\label{eq:thismlr}
M_{LR}\sim\left(\begin{array}{cc}
A_{-1} & \phi_{-4} \\
\phi_{-1} & 0
\end{array}\right)\sim
\left(\begin{array}{ccc} 0 & \epsilon' & \epsilon' \\
-\epsilon' & 0 & 0 \\
0 & \epsilon & 0
\end{array}\right)\langle H_{U}\rangle
\end{equation}
\begin{equation}\label{eq:thismrr}
M_{RR}\sim\left(\begin{array}{cc}
S_{0} & \phi_{-3} \\
\phi_{-3} & 0
\end{array}\right)\sim
\left(\begin{array}{ccc} 0 & 0 & \epsilon' \\
0 & \epsilon & \epsilon \\
\epsilon' & \epsilon &
0
\end{array}\right)\Lambda_{R}
\end{equation}
The U(1) charges in this model allow us to obtain the desired
suppression factors in $Y_U$, without necessitating nontrivial
GUT transformation properties for the flavons, assuming a GUT
is present at all.  While Eqs.~(\ref{eq:thismlr}) and (\ref{eq:thismrr}) 
are of the desired form for neutrino phenomenology, it should be noted 
that this particular model also provides for a large $2$-$3$ mixing angle 
via the diagonalization of $Y_L$.  The numerical analysis for
these textures would therefore be slightly different from that presented 
in Section~3, but the results would be qualitatively unchanged.

\subsection{Discrete Subgroups}

Finally, we mention briefly that any of the models we have discussed  
(and in fact any U(2) model described in the literature~\cite{u2other}) 
can be mapped to an equivalent model based on the discrete group $T'$.  
For example, one possible mapping is
\[
\bf{3} \rightarrow \bf{3} \quad \bf{2}\rightarrow \bf{2}^{-} \quad
\bf{1}\rightarrow \bf{1}^{0} \,\,\, ,
\]
where the notation for $T'$ representations shown on the right is 
explained in Refs.~\cite{acl1,acl2}.  Other mappings exist that render a 
given model free of discrete gauge anomalies~\cite{db}, at least if the
anomalies associated with Abelian factors are canceled via the Green-Schwarz 
mechanism~\cite{gs}.  Thus, realizations of the textures we have studied 
in models with discrete gauge flavor symmetries are also possible.  (For
other approaches to reproducing U(2) physics from discrete groups, see
Ref.~\cite{raby}.)

\section{Conclusions}

In this paper, we have illustrated a simple point:  Models based on 
spontaneously broken non-Abelian symmetries can naturally provide for 
two large neutrino mixing angles, even while quark and charged lepton Yukawa 
textures are hierarchical.  In particular, we have considered U(2)-like 
textures -- textures that can arise in a variety of models that incorporate
the two-step breaking of a non-Abelian symmetry with a subgroup that rotates 
first generation fields by a phase.  We showed how bimaximal mixing could be 
obtained in such models without tuning of parameters, and how perturbations
about these textures, arising in realistic models,  could quantitatively 
explain atmospheric and large-angle solar neutrino oscillations.  Finally,
we presented a number of toy models that reproduce the textures that 
we considered numerically in our model-independent discussion.  While
these models are viable, they nonetheless suggest that better high-energy
realizations are yet to be found. The ideas presented here may therefore
be useful in the eventual formulation of a compelling and comprehensive
theory of fermion masses.


%
\begin{acknowledgments}
C.D.C. and P.M. thank the National Science Foundation for support 
under Grant No.\ PHY-9900657, and the Jeffress Memorial Trust for 
support under Grant No.~J-532. A.A.'s work was supported in part by the 
Department of Energy under grant DE-FG02-91ER40676.
\end{acknowledgments}


\end{document}